\documentclass[twocolumn,showpacs,preprintnumbers,amsmath,amssymb,notitlepage]{revtex4-1}

\usepackage{amsfonts}
\usepackage{amsmath}
\usepackage{amssymb}
\usepackage{graphicx}
\usepackage{dcolumn}
\usepackage{bm}
\usepackage{tabularx}
\usepackage{epstopdf}
\usepackage[dvipsnames]{xcolor}
\usepackage{mathrsfs}
\usepackage{booktabs}

\begin{document}

\title{Entropy-based approach to missing-links prediction}

\author{Federica Parisi}
\affiliation{IMT School for Advanced Studies Lucca, Piazza S.Francesco 19, 55100 Lucca - Italy}
\author{Guido Caldarelli}
\affiliation{IMT School for Advanced Studies Lucca, Piazza S.Francesco 19, 55100 Lucca - Italy}
\author{Tiziano Squartini}
\affiliation{IMT School for Advanced Studies Lucca, Piazza S.Francesco 19, 55100 Lucca - Italy}

\date{\today}

\begin{abstract}
Link-prediction is an active research field within network theory, aiming at uncovering missing connections or predicting the emergence of future relationships from the observed network structure. This paper represents our contribution to the stream of research concerning missing links prediction. Here, we propose an entropy-based method to predict a given percentage of missing links, by identifying them with the most probable non-observed ones. The probability coefficients are computed by solving opportunely defined null-models over the accessible network structure. Upon comparing our likelihood-based, local method with the most popular algorithms over a set of economic, financial and food networks, we find ours to perform best, as pointed out by a number of statistical indicators (e.g. the precision, the area under the ROC curve, etc.). Moreover, the entropy-based formalism adopted in the present paper allows us to straightforwardly extend the link-prediction exercise to \emph{directed} networks as well, thus overcoming one of the main limitations of current algorithms. The higher accuracy achievable by employing these methods - together with their larger flexibility - makes them strong competitors of available link-prediction algorithms.
\end{abstract}


\maketitle

\section*{Introduction}

Link-prediction is an active research field within network theory, aiming at uncovering missing connections (e.g. in incomplete datasets) or predicting the emergence of future relationships from the observed network structure. Loosely speaking, the missing links prediction problem can be stated by asking the following question: \emph{given a snapshot of a network, can the next most-likely links to be established be predicted?} Such an issue is relevant in many research areas, such as social networks \cite{Nowell2003,Pavlov2007,Berlusconi2016,Jalili2017}, protein networks \cite{Barzel2013,Singh2017}, brain networks \cite{Cannistraci2013}, etc.

To this aim, several algorithms have been proposed so far. Overall, ``recipes'' for link-prediction can be classified as belonging to either two main classes, \emph{similarity-based} algorithms or \emph{likelihood-based} algorithms \cite{Lu2011,Zhao2015}. Both classes of algorithms output a list of scores to be assigned to non-observed links: while the similarity-based ones may employ local \cite{Barabasi1999}, quasi-local \cite{Cannistraci2013,Jaccard1901,Sorensen1948,Salton1983,Adamic2003,Zhou2009} or global information \cite{Katz1953,Lu2015,Zhao2015} (e.g. the nodes degree, the degree of common neighbours and the length of paths connecting any two nodes, respectively), the likelihood-based ones \cite{Guimera2009,Tan2014,Pan2016} are defined by a likelihood function whose maximization provides the probability that any two nodes are connected. This is usually achieved by assuming that some kind of benchmark information is known and by treating it as a constraint to account for. An alternative classification distinguishes between algorithms employing purely structural information (either binary or weighted \cite{Lu2011}) and algorithms making use of some kind of external information as well (e.g. nodes attributes \cite{Liao2015}).

This paper represents our contribution to the stream of research concerning missing links prediction. A novel algorithm is proposed, building upon a series of results concerning constrained entropy-maximization \cite{Park2004,Garlaschelli2008,Squartini2011}. In a nutshell, we advance the hypothesis that the tasks of predicting missing links and reconstructing a given network structure share many similarities worth to be further explored. The method we propose in the present work makes a first step in this direction, by employing entropy-based null-models to approach the link-prediction problem. As a last remark, we notice that while the problem of \emph{missing links prediction} is usually associated to the problem of \emph{spurious links identification}, here we only address the former one.

The remainder of the paper is organized as follows. In the ``Methods'' section an overview of the missing links prediction problem is provided, together with a detailed description of the method we propose here. The ``Data'' section contains a synthetic description of the datasets used for testing our methods. In the ``Results'' section, we compare our method with the most common link-prediction algorithms and we comment on the results in the ``Discussion'' section.

\section*{Methods}

In order to fix the formalism, let us briefly reformulate the link-prediction problem \emph{ab initio}.

Let us indicate with the symbol $\mathbf{A}$ the adjacency matrix of the observed network and with the symbol $E$ the corresponding set of \emph{observed links}: as a consequence, upon indicating with $U$ the set of all nodes pairs, $U\setminus E$ will be referred to as to the set of \emph{non-existent links}. In order to fully control a given recipe for link-prediction, the link set is usually partitioned into a \emph{training set}, $E^T$, and a \emph{probe set}, $E^P=E\setminus E^T$. The former is used in the ``calibration'' phase of a given prediction algorithm, while the latter is used for testing it: links belonging to $E^P$ are, in fact, removed, thus constituting the actual ``prediction target''. We denote with $|E^P|\equiv L_{miss}$ the cardinality of the probe set, corresponding to the number of missing links. Naturally, the adjacency matrix is partitioned as well: the portion of it corresponding to the training set will be indicated with the symbol $\mathbf{A}^T$. The union of the missing links set and the non-existent links set $E^{N}=E^P \cup\:U\setminus E\equiv U\setminus E^T$ will be referred to as to the set of \emph{non-observed links}.

Link-prediction algorithms output a list of scores to be assigned to non-observed links. Upon indicating with $i$ and $j$ the nodes constituting the extremes of non-observed links, the most traditional recipes are quickly reviewed below. In what follows, we will focus on the algorithms employing either local or quasi-local information.

\subsection*{Link-prediction for undirected networks}

\begin{itemize}
\item[$\bullet$]The simplest recipe to define scores is based the number of common neighbours (CN) of $i$ and $j$

\begin{equation}
s_{ij}^{CN}=|\Gamma(i)\cap\Gamma(j)|;
\end{equation}

\item[$\bullet$] a slightly more elaborate function of it is represented by the Jaccard coefficient (J), which discounts the information encoded into the size of the nodes neighbourhoods:

\begin{equation}
s_{ij}^{J}=\frac{|\Gamma(i)\cap\Gamma(j)|}{|\Gamma(i)\cup\Gamma(j)|}=\frac{s_{ij}^{CN}}{k_i+k_j-s_{ij}^{CN}};
\end{equation}

\item[$\bullet$]algorithms based on the information provided by nodes degrees exist. The simplest example is provided by the one inspired to the ``preferential attachment'' (PA) mechanism, whose generic score reads

\begin{equation}\label{pau}
s_{ij}^{PA}=k_i\cdot k_j;
\end{equation}

\item[$\bullet$] other, instead, are defined by the inverse of some kind of function of the neighbours degree (according to the original Adamic-Adar - AA - prescription or subsequent variations, as the ``resource allocation'' - RA - one)

\begin{equation}
s_{ij}^{RA}=\sum_{l\in\Gamma(i)\cap\Gamma(j)}\frac{1}{k_l},\:s_{ij}^{AA}=\sum_{l\in\Gamma(i)\cap\Gamma(j)}\frac{1}{\ln k_l};
\end{equation}

\item[$\bullet$] modifications of the aforementioned indices have been recently proposed, encoding information on the link density of the neighbourhood of each pair of nodes. These indices are the so-called CAR-based ones \cite{Cannistraci2013} and prescribe to ``correct'' the scores above by adding a factor $|\gamma(l)|$, counting how many neighbours of node $l\in\Gamma(i)\cap\Gamma(j)$ are also common neighbours of $i$ and $j$. More explicitly

\begin{eqnarray}
s_{ij}^{CAR}&=&s_{ij}^{CN}\cdot\sum_{l\in\Gamma(i)\cap\Gamma(j)}\frac{|\gamma(l)|}{2},\\
s_{ij}^{CJC}&=&\frac{s_{ij}^{CAR}}{|\Gamma(i)\cup\Gamma(j)|},\\
s_{ij}^{CPA}&=&(e_i+s_{ij}^{CAR})\cdot(e_j+s_{ij}^{CAR}),\\
s_{ij}^{CRA}&=&\sum_{l\in\Gamma(i)\cap\Gamma(j)}\frac{\gamma_l}{k_l},\\
s_{ij}^{CAA}&=&\sum_{l\in\Gamma(i)\cap\Gamma(j)}\frac{\gamma_l}{\ln k_l}
\end{eqnarray}
where $e_i$ indicates the external degree of node $i$, i.e. the number of neighbours of $i$ that \emph{are not} neighbours of $j$.
\end{itemize}

\subsection*{Entropy-based approach to link-prediction}

The rationale of our method is based upon the concept of \emph{network reconstructability}. In other words, provided that the \emph{accessible} portion $\mathbf{A}^T$ of a network is satisfactorily reproduced by a given amount of topological information, it is reasonable to suppose that the latter allows the \emph{inaccessible} portion to be inferred with reasonable accuracy as well. Invoking the aforementioned concept allows us to rephrase the link imputation problem within the network reconstruction framework, making it possible to employ the techniques developed there.

From a technical point of view, our algorithm is a local, likelihood-based one. It rests upon the information provided by local, topological quantities, which are enforced as constraints of a maximization procedure defined within the Exponential Random Graph (ERG) framework \cite{Park2004,Squartini2011}. In the case of binary, undirected networks, constraints are represented by nodes degrees, i.e. $\vec{k}(\mathbf{A}^T)$ and the ERG framework leads to the maximization of the likelihood function $\mathcal{L}=\ln P(\mathbf{A}^T)$ where 

\begin{equation}\label{pun}
P(\mathbf{A}^T)=\prod_{i<j}p_{ij}^{a_{ij}}(1-p_{ij})^{1-a_{ij}}
\end{equation}
and $p_{ij}=\frac{x_ix_j}{1+x_ix_j}$. The numerical value of the unknown coefficients $\vec{x}$ is obtained upon solving the system of equations

\begin{equation}\label{sun}
k_i(\mathbf{A}^T)=\sum_{j(\neq i)}p_{ij}=\sum_{j(\neq i)}\frac{x_ix_j}{1+x_ix_j}\:\forall\:i
\end{equation}
(see the appendix for the derivation of the condition above). Our algorithm, which is trained on $\mathbf{A}^T$, prescribes to interpret the probability coefficients $\{p_{ij}\}_{ij\in E^N}$ assigned to the non-observed links, as scores to carry out the link-prediction: upon sorting the coefficients $\{p_{ij}\}_{ij\in E^N}$ in decreasing order, the first $L_{miss}$ largest ones are naturally interpreted as pointing out the $L_{miss}$ \emph{most probable missing links} (notice that such a prescription is based on the assumption that the number of missing links is known, although their identity is not: as a consequence, this number is retained). In other words, the reconstructability assumption underlying our method leads us to interpret the non-observed links which have been assigned the largest probability coefficients as the ones that are most likely to appear given the chosen constraints.

Our recipe has a remarkable, equivalent formulation. In fact, the subset $\mathbf{\Sigma}^*$ of $L_{miss}$ links characterized by the largest probability coefficients identifies the subgraph satisfying the relationship

\begin{equation}\label{sigma}
\mathbf{\Sigma}^*=\substack{\text{\normalsize argmax}\\\mathbf{\Sigma}:|E|(\mathbf{\Sigma})=L_{miss}}\:P(\mathbf{\Sigma}|\mathbf{A}^T)
\end{equation}
with $P(\mathbf{\Sigma}|\mathbf{A}^T)=\prod_{\substack{i<j\\ij\in E^N}}p_{ij}^{\sigma_{ij}}(1-p_{ij})^{\sigma_{ij}}$. Since the maximum value of such a product is achieved once the $L_{miss}$  largest factors are selected, the generic entry $\sigma_{ij}^*$ obeys the following rule: $\sigma_{ij}^*=1$ if $ij$ belongs to the set of $L_{miss}$ most probable missing links and $\sigma_{ij}^*=0$ otherwise; in other words, $\mathbf{\Sigma}^*$ is the subgraph with largest probability among the ones with precisely $L_{miss}$ links. In the remainder of the paper, this approach will be named after the null-model employed to calculate the link scores, i.e. UBCM (Undirected Binary Configuration Model) \cite{Squartini2011}.

\subsection*{Link-prediction for directed networks}

Remarkably, our algorithm can be generalized to approach the missing links prediction problem in directed networks as well. It is enough to maximize the likelihood $\mathcal{L}=\ln P(\mathbf{A}^T)$ where, now, $P(\mathbf{A}^T)=\prod_{i\neq j}p_{ij}^{a_{ij}}(1-p_{ij})^{1-a_{ij}}$ by solving the system of equations

\begin{equation}\label{sdi}
\begin{cases}
k_i^{out}(\mathbf{A}^T)&=\:\:\:\:\sum_{j(\neq i)}p_{ij}=\sum_{j(\neq i)}\frac{x_iy_j}{1+x_iy_j}\:\forall\:i\\
k_i^{in}(\mathbf{A}^T)&=\:\:\:\:\sum_{j(\neq i)}p_{ji}=\sum_{j(\neq i)}\frac{x_jy_i}{1+x_jy_i}\:\forall\:i
\end{cases}
\end{equation}
and consider the coefficients $\{p_{ij}\}_{ij\in E^N}$ as scores to be assigned to the non-observed links (see the appendix for the derivation of the condition above). The proper prediction step is still carried out by applying the recipe defined by eq. \ref{sigma}, with the only difference that, now, the product runs over the \emph{directed} pairs of nodes. In the remainder of the paper, this approach will be named after the null-model employed to calculate the link scores, i.e. DBCM (Directed Binary Configuration Model) \cite{Squartini2011}.\\

Notice, instead, that no unambiguous ways to generalize traditional scores exist. Here we have adopted the (directed) extensions listed below, with the aim of accounting for link directionality whenever possible:

\begin{itemize}
\item[$\bullet$] when considering directed networks, the concept of common neighbours can be replaced by the concepts of ``successors'' and ``predecessors'', i.e. the nodes respectively ``pointed by'' and ``pointing to'' a given node. Upon indicating the set of ``successors'' of $i$ with $\Gamma_S$ and the set of ``predecessors'' of $j$ with $\Gamma_P$, the CN index can be generalized as follows

\begin{equation}
s_{ij}^{CN}=|\Gamma_S(i)\cap\Gamma_P(j)|;
\end{equation}

\item[$\bullet$]building upon the directed version of the CN index, the J index reads

\begin{equation}
s_{ij}^{J}=\frac{|\Gamma_S(i)\cap\Gamma_P(j)|}{|\Gamma_S(i)\cup\Gamma_P(j)|}=\frac{s_{ij}^{CN}}{k_i^{out}+k_j^{in}-s_{ij}^{CN}};
\end{equation}

\item[$\bullet$]the RA and AA indices can be straightforwardly generalized as follows:

\begin{equation}
s_{ij}^{RA}=\sum_{l\in\Gamma(i)\cap\Gamma(j)}\frac{1}{k_l^{tot}},\:s_{ij}^{AA}=\sum_{l\in\Gamma(i)\cap\Gamma(j)}\frac{1}{\ln k_l^{tot}}
\end{equation}
with $k_i^{tot}=k_i^{out}+k_i^{in}$;

\item[$\bullet$]the PA score admits two different generalizations: one employing the total degree of nodes

\begin{equation}\label{pa1}
s_{ij}^{PA'_I}=k_i^{tot}\cdot k_j^{tot}
\end{equation}
and the other employing the nodes out- and in-degree

\begin{equation}\label{pa1}
s_{ij}^{PA'_{II}}=k_i^{out}\cdot k_j^{in};
\end{equation}

\item[$\bullet$]while the CAR-based indices are not straightforwardly generalizable to the directed case, other scores exist aiming at extending the concept of ``closed triad'' to account for link directionality \cite{Schall2014}:

\begin{equation}
s_{ij}^{TC}=\sum_{l\in\Gamma(i)\cap\Gamma(j)}w_{i,j,l}\cdot w(l);
\end{equation}
here, the ``triad weight'' $w_{i,j,l}=\frac{\#T_{i\rightarrow j,l}+\#T_{i\leftrightarrow j,l}}{\#T_{i,j,l}}$ is defined by the (global) number $\#T_{i,j,l}$ of observed, \emph{open} triads of the particular kind $T_{i,j,l}$, the (global) number $\#T_{i\rightarrow j,l}$ of observed, \emph{closed} triads via a directed link from $i$ to $j$ and the (global) number $\#T_{i\leftrightarrow j,l}$ of observed, \emph{closed} triads via a reciprocal link between $i$ and $j$; $w(l)$ is, instead, a node-specific weight that can be set either to $w(l)=\frac{1}{k_l}$ or to 1. In order to avoid misinterpretations, we set the weight to 1.
\end{itemize}

\subsection*{Testing link-prediction}

Once a link-prediction algorithm has been defined, a number of statistical indices exist to test its effectiveness. In what follows we will briefly review the ones we have employed in the present paper to compare the aforementioned algorithms. The first index we have considered is the \emph{true positive rate} (also known with the name of \emph{precision}), defined as 

\begin{equation}
\text{TPR}=\frac{L_r}{L_{miss}}
\end{equation}
and quantifying the percentage of missing links that are correctly recovered (i.e. the number $L_r$ of rightly identified missing links within the list of the first $L_{miss}$ links with the largest score). A similar-in-spirit index is the \emph{accuracy}

\begin{equation}
\text{ACC}=\frac{L_r+L_{ne}}{|E^N|},
\end{equation}
quantifying the percentage of correctly classified links (i.e. both the missing ones and the non-existent ones) with respect to the total number of non-observed links. The third index we consider is the traditional \emph{area under the ROC curve}, or AUC, proxied by the number 

\begin{equation}
\text{AUC}=\frac{n'+n''/2}{n};
\end{equation}
$n'$ counts the number of times a missing-link is assigned a higher probability than to a non-existent one, while $n''$ accounts for the number of times they are assigned an equal probability. The denominator $n$ coincides with the total number of comparisons (i.e. the number of missing links times the number of non-existent links). This index is intended to quantify the probability that any missing-link is assigned a score that is larger than the score assigned to any non-observed link. If all scores were i.i.d. the AUC value should be distributed around an expected value of $1/2$: therefore, the extent to which the AUC value exceeds 0.5 provides an indication of how much better the algorithm performs than pure chance.\\

The set of missing links is usually randomly removed: we have followed such a procedure, by 1) randomly removing the 10\% of links 10 times, 2) quantifying the performance of the algorithms above, by computing the three aforementioned indices over each sample, 3) averaging these values over the sample set (the sample standard deviation is used to proxy the estimation error.)

\section*{Data}

\begin{table}
\begin{tabular}{@{\extracolsep{5pt}}lc|c|c} 
\hline
\multicolumn{4}{c}{\bf World Trade Web}\\
\hline
\text{\bf BUN} & $N$ & $L$ & $\langle k\rangle$ \\
\hline
& 145 [85,187] & 5901 [1678,10254] & 76.3 [39.5,109.7]\\
\hline
\text{\bf BDN} & & $L$ & $\langle k\rangle$ \\
\hline
& & 10604 [2871,20107] & 68.2 [33.8,107.5]\\
\hline
\hline
\multicolumn{4}{c}{\bf e-MID}\\
\hline
\text{\bf BUN} & $N$ & $L$ & $\langle k\rangle$ \\
\hline
& 152 [107,170] & 1755 [742,2333] & 22.5 [13.4,28.6]\\
\hline
\text{\bf BDN} & & $L$ & $\langle k\rangle$ \\
\hline
& & 1827 [754,2477] & 11.7 [6.8,14.9]\\
\hline
\hline
\multicolumn{4}{c}{\bf Dutch Interbank Network}\\
\hline
\text{\bf BUN} & $N$ & $L$ & $\langle k\rangle$ \\
\hline
 & 99 [92,104] & 675 [444,1025] & 13.6 [8.9,20.1]\\
\hline
\text{\bf BDN} & & $L$ & $\langle k\rangle$ \\
\hline
& & 773 [512,1207] & 7.8 [5.2,11.8]\\
\hline
\end{tabular}
\caption{Network statistics for both the undirected (BUN) and directed (BDN) version of the World Trade Web, e-MID and the Dutch Interbank Network: for each quantity (number of nodes $N$, number of links $L$ and average degree $\langle k\rangle$) the mean value across the temporal snapshots and the range are reported.}
\label{tab1}
\end{table}

Our approach to link-prediction has been tested on a number of economic and financial datasets (see table \ref{tab1}) and on several food-webs (see table \ref{tab2}).

As a first dataset, we have considered the World Trade Web (WTW) across a period of 51 years, i.e. from 1950 to 2000. The dataset in \cite{Gleditsch2002} collects yearly, bilateral, aggregated data on exports and imports (the generic entry $m_{ij}^{agg}(y)$ is the sum of the single commodity-specific trade exchanges between $i$ and $j$ during the year $y$). The binary, directed representation of the WTW we have considered here has been obtained by linking any two nodes whenever the corresponding element $m_{ij}^{agg}(y)$ is strictly positive, i.e. $a_{ij}(y)=\Theta[m_{ij}^{agg}(y)]$.

As a second dataset, we have considered the Dutch Interbank Network (DIN) across a period of 11 years, i.e. from 1998 to 2008 \cite{tVeld2014}. Such a dataset collects quarterly data on exposures between Dutch banks, larger than 1.5 million euros and with maturity shorter than one year.

As a third dataset, we have considered the e-MID (i.e. the electronic Market for Interbank Deposits) network in a series of 61 temporal snapshots, corresponding to the maintenance periods (and ranging from 2005 to 2010). In this case, links represent granted loans \cite{Iori2006}. As for the WTW, the binary, directed representations of both the DIN and e-MID have been obtained by linking any two nodes whenever a positive weight is observed between them.

The three real-world systems above are defined by directed connections. In order to evaluate the performance of the link-prediction algorithms considered in the present paper on undirected networks, we have properly symmetrized the adjacency matrices of these systems, according to the prescription $a_{ij}^{sym}=a_{ij}+a_{ji}-a_{ij}a_{ji}$.

\begin{table}
\begin{tabular}{@{\extracolsep{5pt}}lcccc} 
\hline
\textbf{Food-webs} & $N$ & $L$ & $\langle k\rangle$ \\
\hline
Chesapeake Bay & 39 & 177 & 4.54\\
Lower Chesapeake Bay & 37 & 178 & 4.81\\
Middle Chesapeake Bay & 37 & 209 & 5.65\\
Upper Chesapeake Bay & 37 & 215 & 5.81\\
Everglades Marshes & 69 & 916 & 13.28\\
Florida Bay & 128 & 2106 & 16.45\\
Grassland & 88 & 137 & 1.56\\
Little Rock Lake & 183 & 2494 & 13.63\\
Maspalomas Lagoon & 24 & 82 & 3.42\\
Michigan Lake & 39 & 221 & 5.67\\
Mondego Estuary & 46 & 400 & 8.70\\
Narragansett Bay & 35 & 220 & 6.29\\
Rhode River Watershed & 19 & 53 & 2.79\\
Silwood Park & 154 & 370 & 2.40\\
St Marks River & 54 & 356 & 6.59\\
St Marks Seagrass & 49 & 226 &  4.61\\
St Martin Island & 45 & 224 & 4.98\\ 
Ythan Estuary & 135 & 601 & 4.45\\
\hline
\end{tabular}
\caption{Network statistics (number of nodes $N$, number of links $L$ and average degree $\langle k\rangle$ for the directed version of 18 different food-webs \cite{Squartini2011}.}
\label{tab2}
\end{table}

Food-webs, instead, are considered in their binary, directed version only: if species $i$ preys on species $j$, a directed link is drawn from $j$ to $i$.

\section*{Results}

\begin{figure*}[t!]
\includegraphics[width=0.85\textwidth]{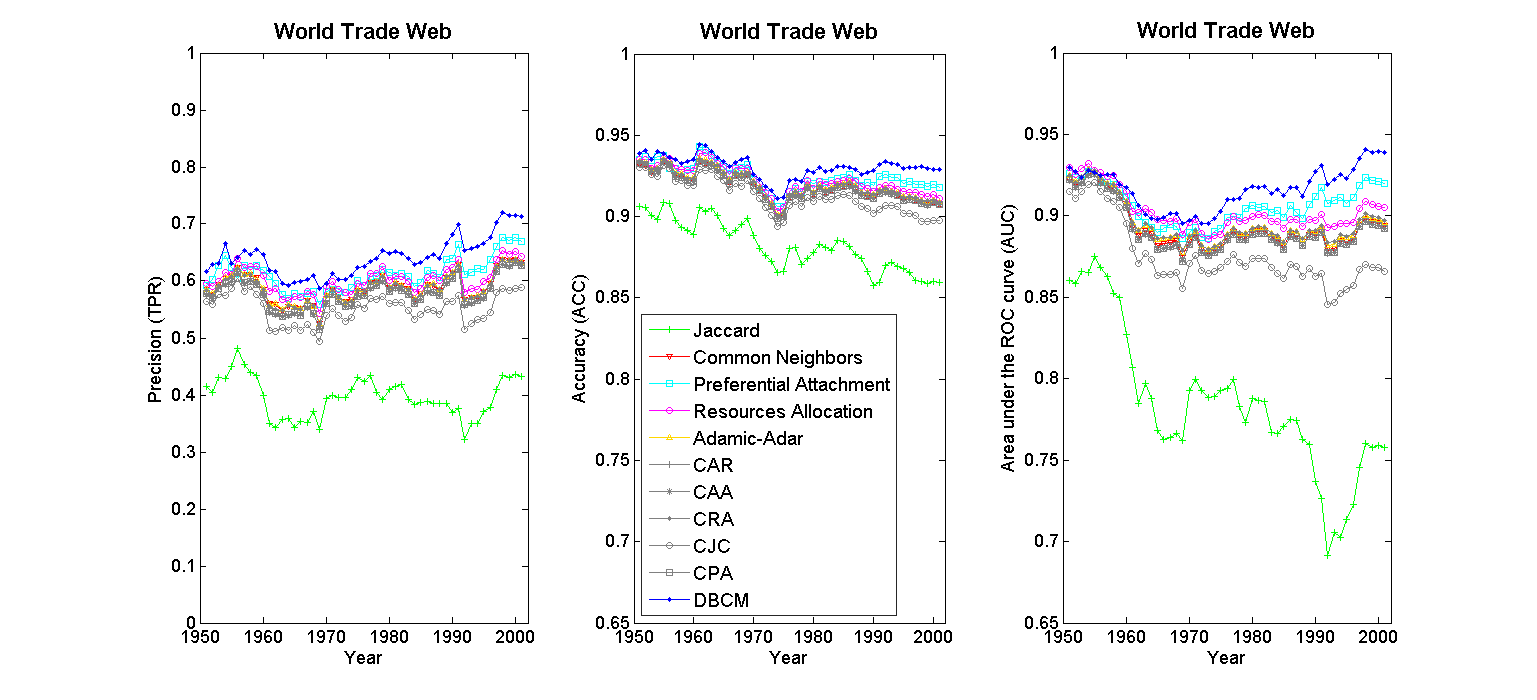}
\includegraphics[width=0.85\textwidth]{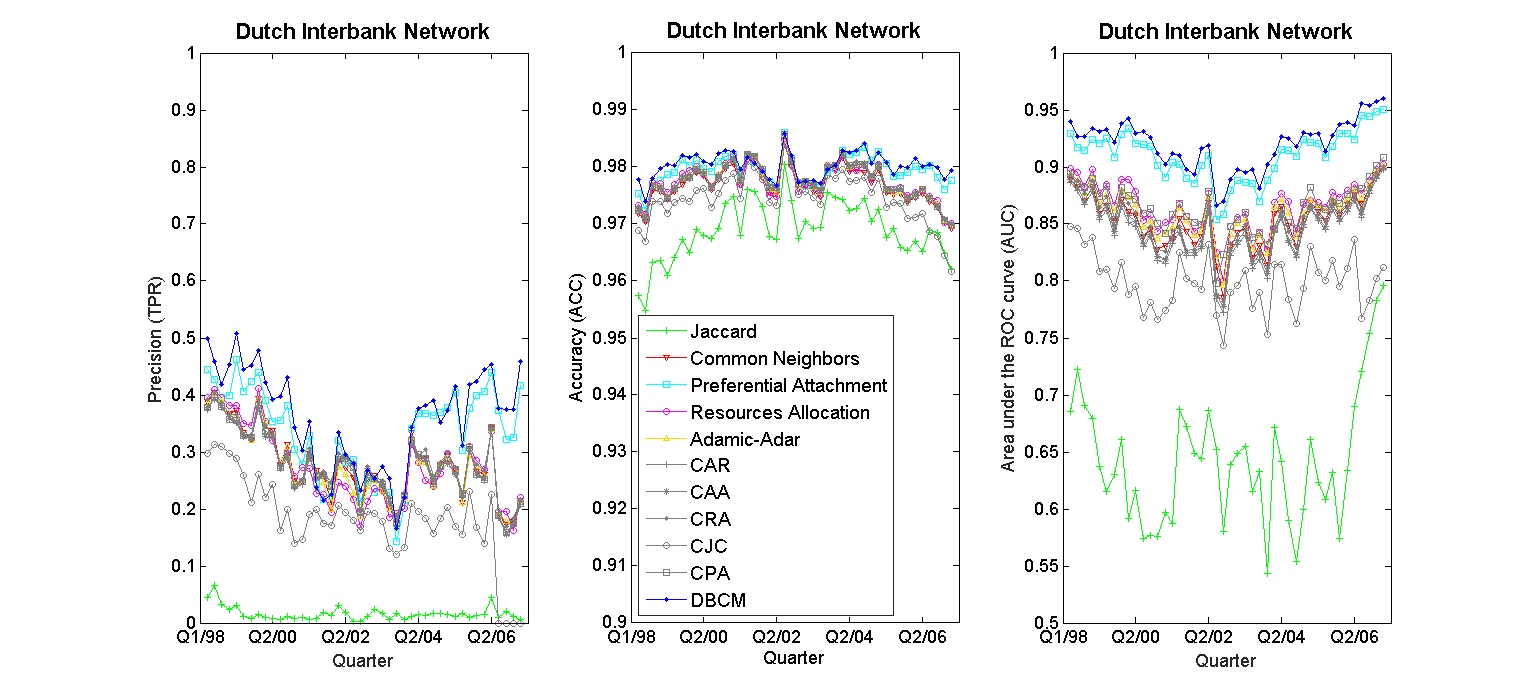}
\includegraphics[width=0.85\textwidth]{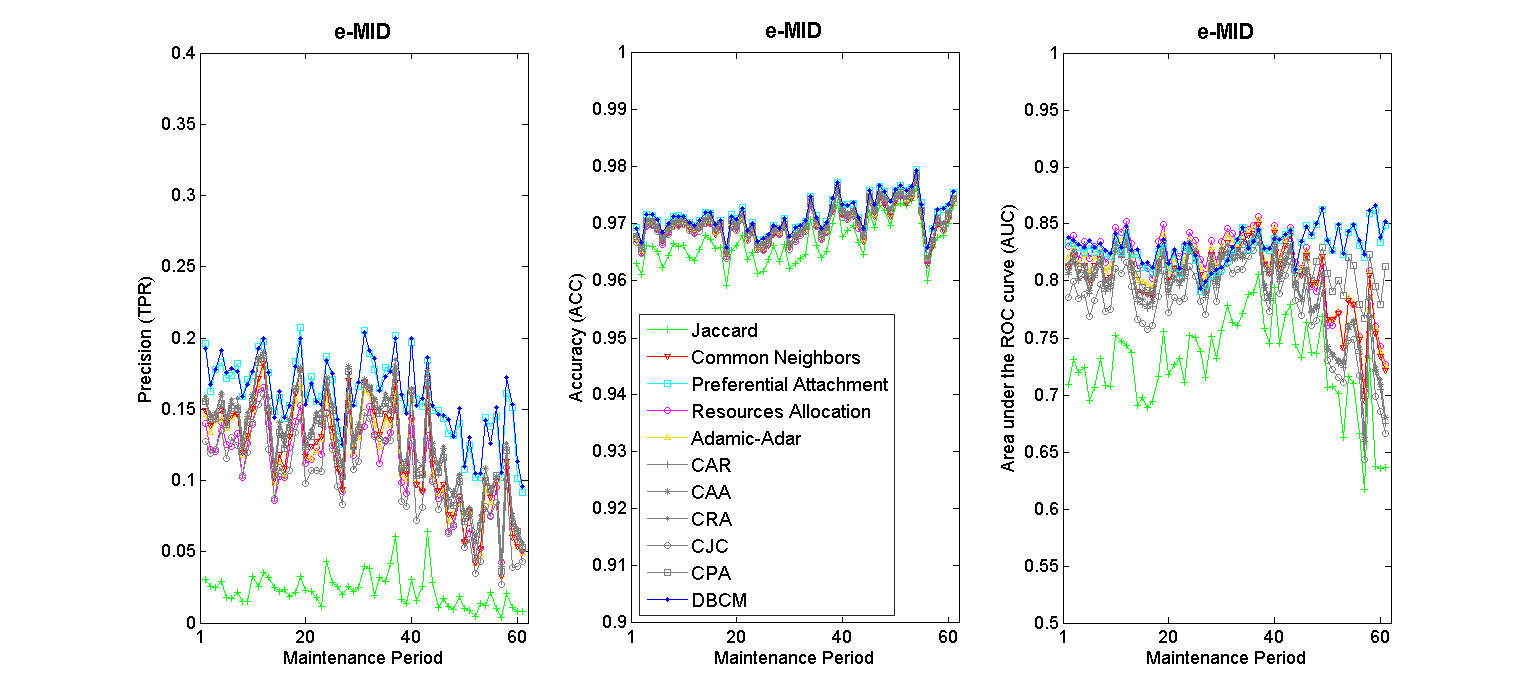}
\caption{Comparison of link-prediction algorithms when applied to the undirected version of the WTW (top), the DIN (middle) and e-MID (bottom). Left panel: evolution of precision; middle panel: evolution of accuracy; right panel: evolution of AUC. Our entropy-based approach to link-prediction performs better than other algorithms, across the vast majority of temporal snapshots. Notice that PA implements the recipe $s_{ij}^{PA}=k_i\cdot k_j$. See also fig. \ref{figA1} for a visual inspection of the errors accompanying the estimations shown here.}
\label{fig1}
\end{figure*}

\begin{figure*}[t!]
\includegraphics[width=0.87\textwidth]{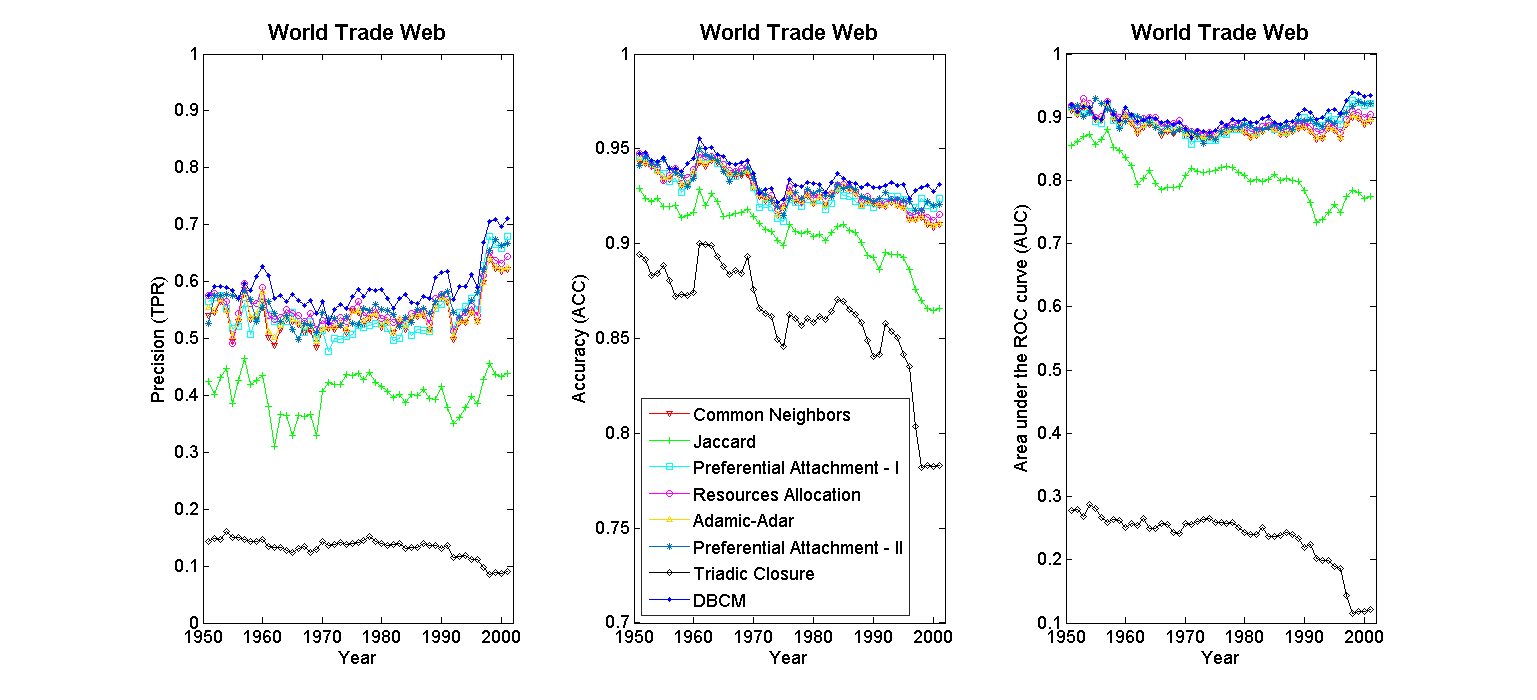}
\includegraphics[width=0.87\textwidth]{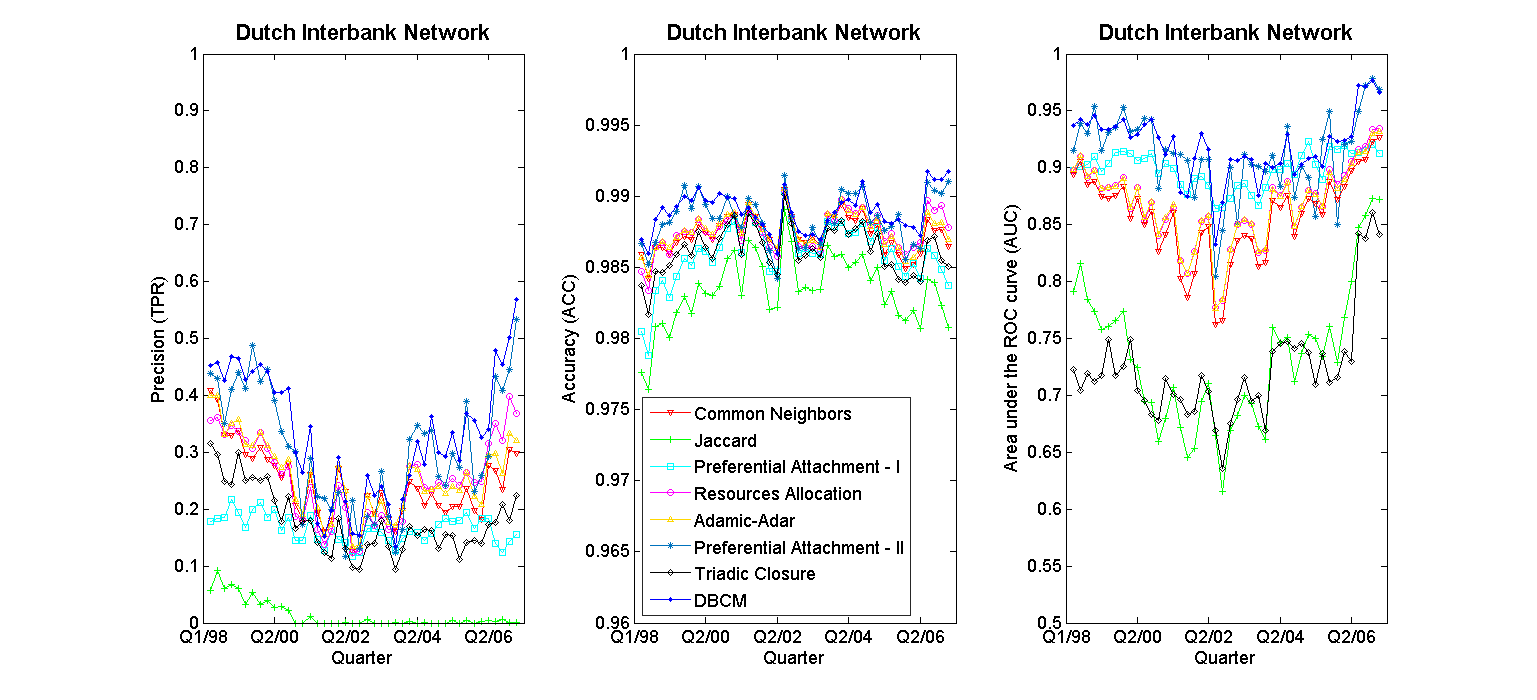}
\includegraphics[width=0.87\textwidth]{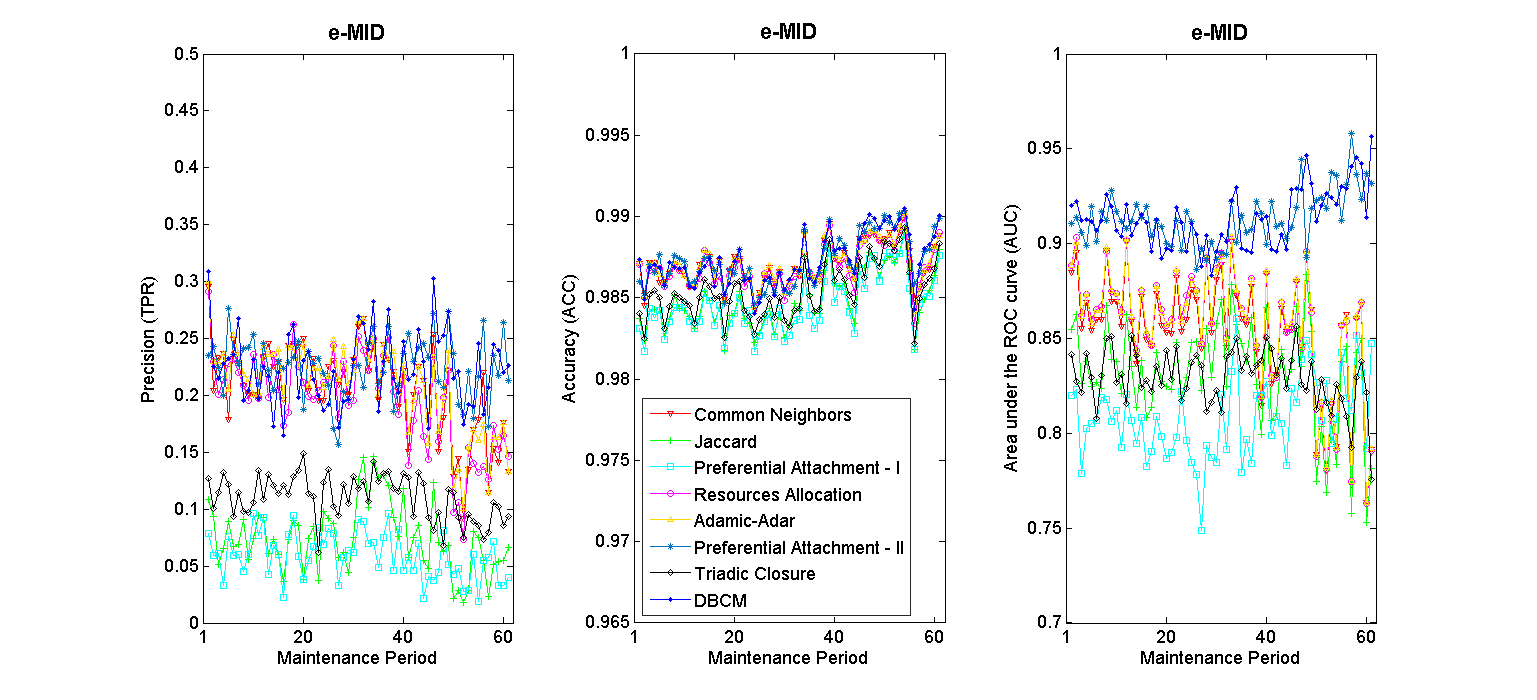}
\caption{Comparison of link-prediction algorithms when applied to the directed version of the WTW (top), the DIN (middle) and e-MID (bottom). Left panel: evolution of precision; middle panel: evolution of accuracy; right panel: evolution of AUC. Our entropy-based approach to link-prediction performs better than other algorithms, across the vast majority of temporal snapshots. Notice that while PA-I implements the recipe $s_{ij}^{PA'}=k_i^{tot}\cdot k_j^{tot}$, PA-II implements the recipe $s_{ij}^{PA''}\propto k_i^{out}\cdot k_j^{in}$.}
\label{fig2}
\end{figure*}

\begin{figure*}[t!]
\includegraphics[width=\textwidth]{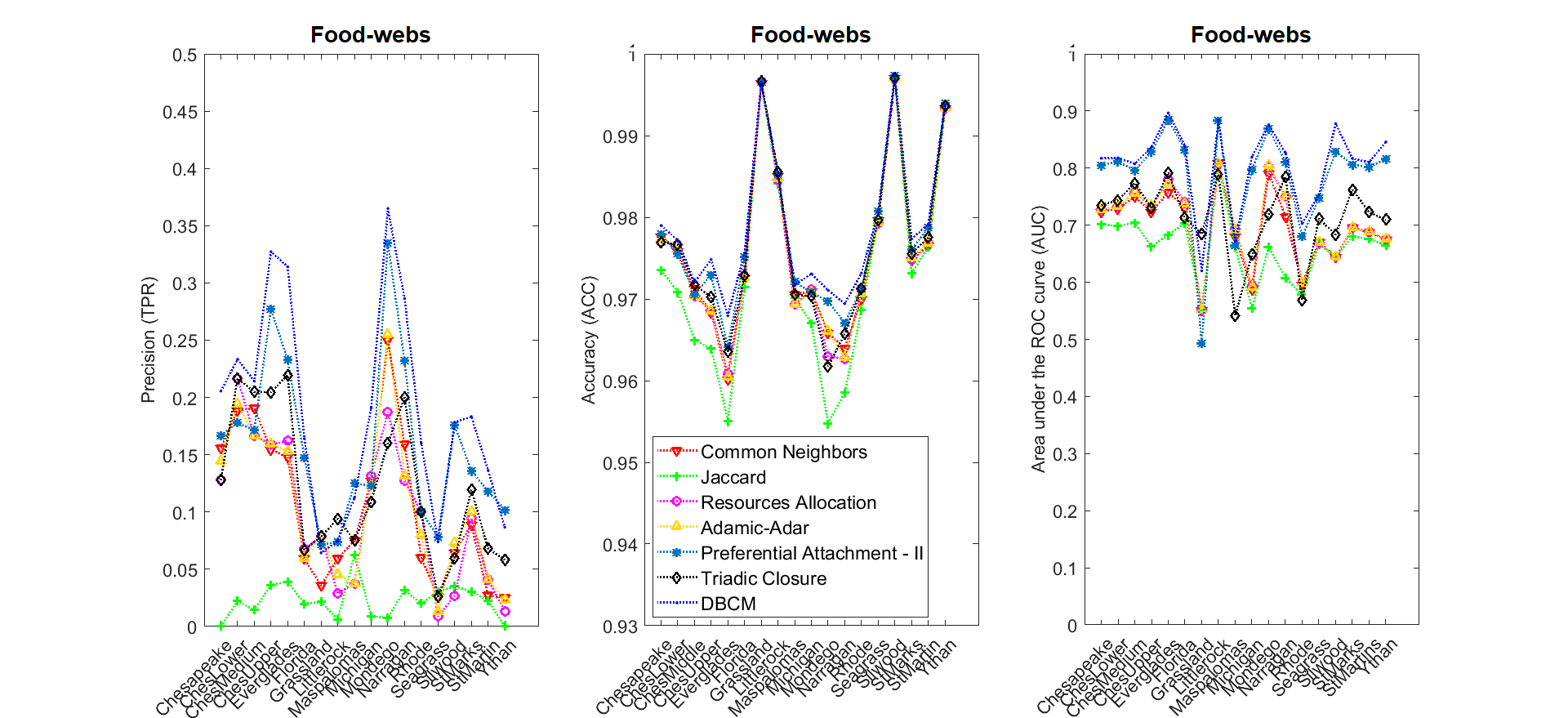}
\caption{Comparison of link-prediction algorithms when applied to the (directed version) of the food-webs. Our entropy-based approach to link-prediction performs better than the other algorithms. Here, we have retained only the recipe $s_{ij}^{PA''}\propto k_i^{out}\cdot k_j^{in}$.}
\label{fig3}
\end{figure*}

The performance of our link-prediction algorithm is shown in figs. \ref{fig1}, \ref{fig2} and \ref{fig3}: the three panels of fig. \ref{fig1} and \ref{fig2} refer to the WTW, the DIN and e-MID respectively while food-webs are reported in fig. \ref{fig3}.

As a general comment, our method performs better than the other algorithms, with respect to all considered indices. The success of the method is particularly evident when considering the AUC index, proxying the probability of (correctly) assigning a larger score to a missing-link than to a non-existent link.

We argue the success of our algorithm to rest upon a core result that has been verified in a number of previous works \cite{Squartini2011,Squartini2011b,Cimini2015}: the (purely) topological structure of the networks considered here can be reconstructed, to a large degree of accuracy, by enforcing the information encoded into the degree sequences alone; very likely, thus, the same amount of information also defines an accurate recipe to spot potential missing links. Otherwise stated, the level of ``complexity'' of the considered networks seems to be largely encoded into the degree sequences, thus requiring (just) their enforcement to be fully accounted for.

The founding principle of our approach is, thus, radically different from the one inspiring other link-prediction algorithms: we aim at finding the (most likely) generative process for the network at hand, while other methods define increasingly detailed procedures with little control on the ``quality'' of the included information. This becomes evident when considering that other algorithms (i.e. the CN, J, RA, AA and the CAR-based ones) employ a larger amount of information than the UBCM- and DBCM-based ones: while the latter take as input just the nodes degrees, the former exploit the information provided by the whole set of common neighbours. This may indicate that the information encoded into the neighbourhood of any two nodes - supposedly providing more information than the one encoded into the degrees alone - is, actually, a mere consequence of lower-order statistics (i.e. the degrees themselves).

This also sheds light on the reason why our algorithm is less sensitive than others to the original value of link density: provided that our entropy-based recipes successfully individuate the process generating the networks at hand, the number of observed links is automatically accounted for.

Our comparison also points out that one of the factors determining the goodness of a given link-prediction algorithm concerns \emph{how} the available information is used. An illustrative example is provided by the performance of the PA algorithm defined by eq. \ref{pa1}, requiring the same basic knowledge of our entropy-based recipes, i.e. the degree sequences of nodes. As clear upon inspecting the e-MID directed case, the assumption that any two nodes establish a connection with a probability that is proportional to their \emph{total} degrees fails to capture the process shaping the network structure; entropy-maximization, on the other hand, makes a better use of the available information, by retaining the information on link directionality (that indeed plays a role, completely ignored by the aforementioned PA prescription).

Interestingly, the DBCM recipe described in the ``Methods'' section induces the ``correct'', directed generalization of the PA algorithm, defined by eq. \ref{pa2} and outperforming the one defined by eq. \ref{pa1}. For sparse networks, in fact, the DBCM probability coefficients can be approximated as follows

\begin{equation}\label{pa2}
p_{ij}^{\text{DBCM}}\simeq\frac{k_i^{out}\cdot k_j^{in}}{L}
\end{equation}
a simplified prescription that performs very similarly to the entropy-based algorithm on the DIN and e-MID; on dense networks - as the WTW - the DBCM performs much better, instead. The UBCM, on the other hand, reduces to $p_{ij}^{\text{UBCM}}\propto k_i\cdot k_j$, i.e. the undirected PA prescription defined by eq. \ref{pau}.

Finally, let us comment on the performance of the TC index. The algorithm employing it has been designed to provide a solution to the problem of forecasting new connections among social networks users. By definition, it only predicts new links among \emph{disconnected} nodes, disregarding all nodes pairs connected by, e.g. a non-reciprocated link. This explains the poor performance of the algorithm in our context, despite it performs satisfactorily to solve the specific task it was designed for \cite{Schall2014}.

\section*{Discussion}

Whenever judging the performance of a given link-prediction algorithm, one should consider both the amount of information it requires and the way in which this is employed to carry out the prediction step. While the usual link-prediction algorithms assume the existence of some node-specific tendency at a microscopic level (e.g. social agents tend to close triads), ours focuses on the most likely process that may have generated the considered network. The guessed process is, first, trained on the visible portion of the network and, then, employed to infer the (supposedly) unknown portion of the network: the ``homogeneity'' assumption underlying the whole procedure leads us to expect that a model satisfactorily reproducing the accessible part of a system is also effective in spotting potential missing links.

One of the most effective recipes to tune generative processes is the one based on the entropy-maximization: beside guaranteeing that the available information is encoded in the least-biased way, the ERG framework is also very flexible, being applicable to both undirected and directed networks; other algorithms, on the contrary, rest upon concepts unambiguously defined only for undirected networks (an example is provided by the whole family of CAR indicators, whose core concept - i.e. the ``local community links'' factor $|\gamma (l)|$ - does not admit a straightforward generalization).

Although every newly-proposed algorithm fosters the idea to be applicable to different kinds of systems, the effectiveness of a given (null) model depends on the particular system at hand: while economic, financial and food networks seem to be largely explained by the degree sequences, other systems may require a different (or additional) kind of information.

The results obtained so far on undirected, as well as directed, binary networks push us to look for further extensions of the proposed link-prediction technique. Interesting perspectives are represented by bipartite and weighted networks, for which the link- and weight-imputation topics are still little explored.

\appendix

\begin{figure*}[t!]
\includegraphics[width=0.87\textwidth]{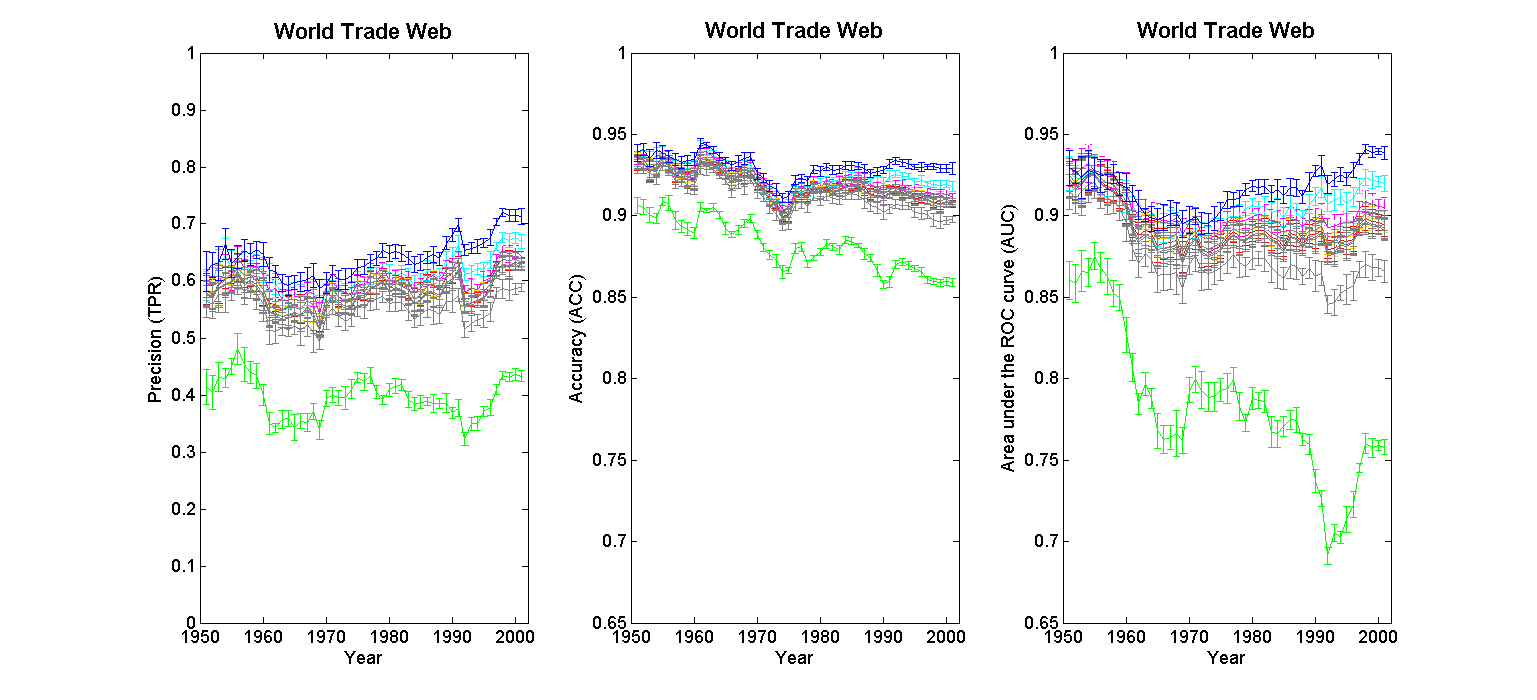}
\includegraphics[width=0.87\textwidth]{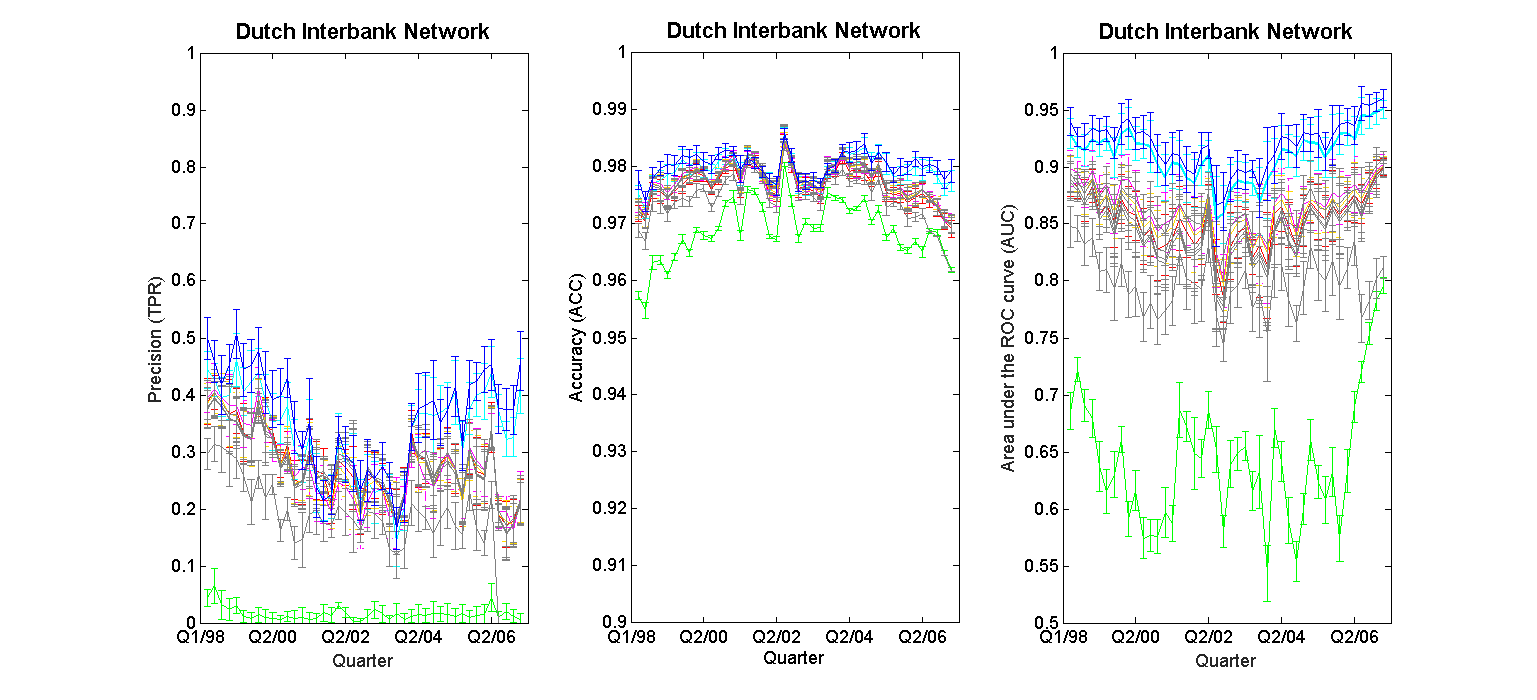}
\includegraphics[width=0.87\textwidth]{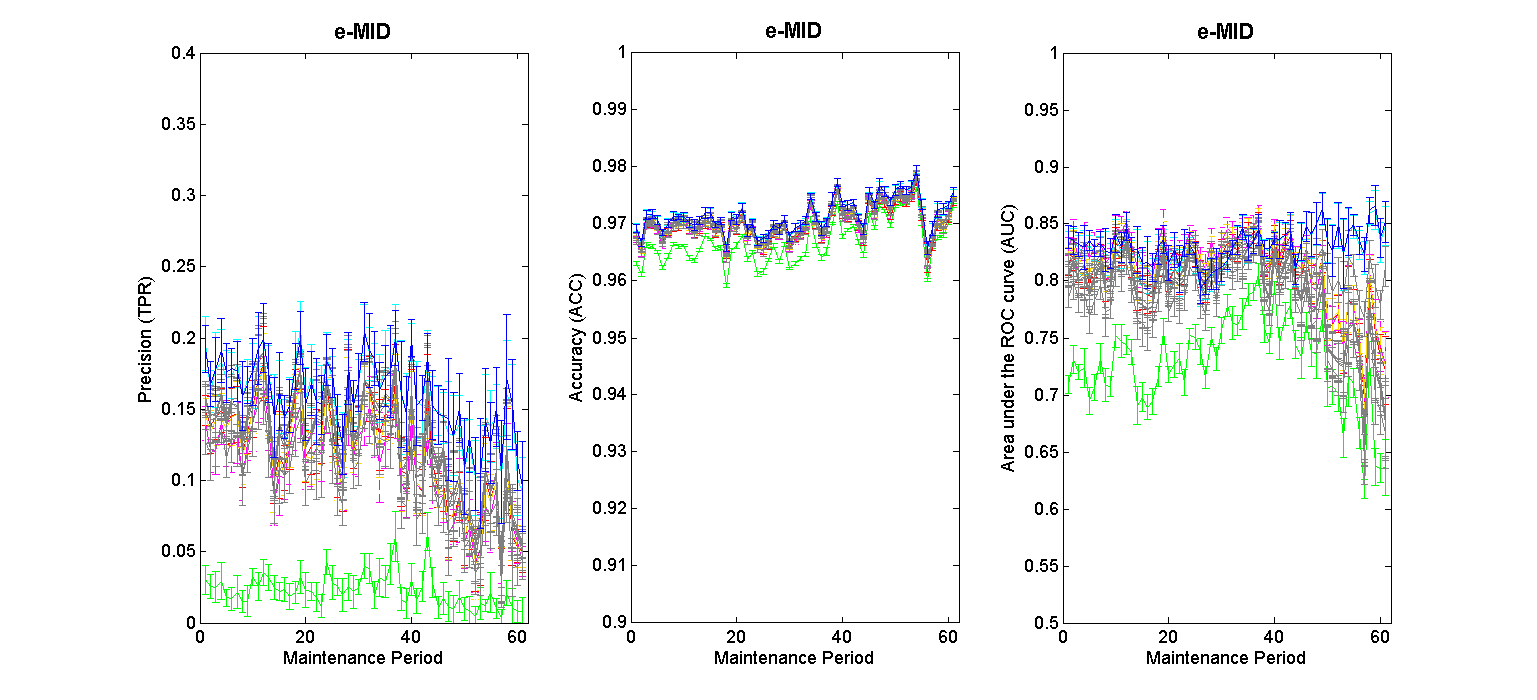}
\caption{Error bars accompanying the estimation of the link-prediction algorithms shown in fig. \ref{fig1}. Panels refer to the undirected version of the WTW (top), the DIN (middle) and e-MID (bottom).}
\label{figA1}
\end{figure*}

\section*{Appendix: Configuration Models}

This Appendix is devoted to the explicit derivation of the undirected and directed version of the Binary Configuration Model. Let us start by defining the core quantity of our approach, i.e. Shannon entropy

\begin{equation}
S[P] =-\sum_{\mathbf{A}\in\mathcal{A}}P(\mathbf{A})\ln P(\mathbf{A});
\end{equation}
representing a functional of the probability distribution $\{P(\mathbf{A})\}_{\mathbf{A}\in\mathcal{A}}$ defined over the ensemble of configurations $\mathcal{A}$. Its constrained maximization represents an inference procedure which has been proved to be maximally non-committal with respect to the missing information \cite{Squartini2011}. To this aim, let us define the Lagrangean function

\begin{equation}
\mathscr{L}[P]=S[P]-\sum_{m=0}^M\theta_m\left(\sum_ {\mathbf{A}\in\mathcal{A}}P(\mathbf{A})C_m(\mathbf{A})-\langle C_m\rangle\right)
\end{equation}
with $C_m$ representing the $m$-th constraint and the $C_0=\langle C_0\rangle=1$ summing up the normalization condition. Upon solving the equation $\frac{\delta\mathscr{L}[P]}{\delta P(\mathbf{A})}=0$ one finds the expression $P(\mathbf{A})=e^{-1-\vec{\theta}\cdot\vec{C}(\mathbf{A})}$ that can be further re-written as

\begin{equation}
P(\mathbf{A})=\frac{e^{-\sum_{m=1}^M\theta_m C_m(\mathbf{A})}}{Z(\vec{\theta})}
\end{equation}
a formula defining the Exponential Random Graph formalism in its full generality. Since we are interested in defining a link-prediction algorithm employing only local information, let us now enforce the nodes degrees as constraints. In the undirected case, this amounts at posing $\vec{C}(\mathbf{A}^T)=\vec{k}(\mathbf{A}^T)$ which leads to the equivalence $\sum_{m=1}^M\theta_mC_m(\mathbf{A}^T)=\sum_{i<j}(\theta_i+\theta_j)a_{ij}(\mathbf{A}^T)$ and, upon identifying $x_i\equiv e^{-\theta_i}$, further leads to eq. \ref{pun}. In the directed case, instead, $\vec{C}(\mathbf{A}^T)=\{\vec{k}^{out}(\mathbf{A}^T), \vec{k}^{in}(\mathbf{A}^T)\}$, leading to $\sum_{m=1}^M\theta_mC_m(\mathbf{A}^T)=\sum_{i\neq j}(\theta_i+\lambda_j)a_{ij}(\mathbf{A}^T)$ and to the directed version of $P(\mathbf{A})$ (a function, now, of $x_i\equiv e^{-\theta_i}$ and $y_i=e^{-\lambda_i}$).

The recipe to estimate the unknown parameters comes from another principle, i.e. the likelihood maximization one. Upon maximizing the function

\begin{equation}
\mathcal{L}=\ln P(\mathbf{A}^T)
\end{equation}
with respect to the unknowns (i.e. $\vec{x}$ in the undirected case and $\{\vec{x}, \vec{y}\}$ in the directed case) the systems of equations \ref{sun} and \ref{sdi} are recovered.

As a last remark, we stress that the computational complexity of the whole algorithm is the one required for solving the systems of equations \ref{sun} and \ref{sdi}. In both cases, the formulation provided in the present paper can be further simplified by limiting ourselves to consider only the distinct values of the degrees \cite{Garlaschelli2008}. This induces the resolution of a reduced system of equations, further lowering the computational complexity of the whole algorithm.

\section*{Acknowledgements}

This work was supported by the EU projects CoeGSS (grant num. 676547), DOLFINS (grant num. 640772), MULTIPLEX (grant num. 317532), Shakermaker (grant num. 687941), SoBigData (grant num. 654024).

\section*{Authors Contributions}

F. Parisi and T. Squartini developed the method. F. Parisi performed the analysis. F. Parisi, G. Caldarelli and T. Squartini wrote the manuscript. All authors reviewed and approved the manuscript.

\end{document}